\documentclass[reprint,amsmath,amssymb,aps,prb,]{revtex4-2}

\usepackage{graphicx}
\usepackage{amsmath}
\usepackage{xcolor}
\usepackage{dcolumn}
\usepackage{enumitem}
\usepackage{epstopdf}

\begin{document}

\title{Magnetic resonances in EuSn$_2$As$_2$ single crystal.}

\author{I.~A.~Golovchanskiy$^{1,2,3,*}$, E.~I.~Maltsev$^{1,4}$, I.V.~Shchetinin$^2$, V.A.~Vlasenko$^4$, P.S.~Dzhumaev$^{2,5}$, K.~Pervakov$^4$, O.V.~Emelyanova$^{5,6}$, A.Yu.~Tsvetkov$^4$, N.N.~Abramov$^2$, V.M.~Pudalov$^{4,7}$, V.S.~Stolyarov$^{1,2,3}$}

\affiliation{
$^1$ Moscow Institute of Physics and Technology, State University, 9 Institutskiy per., Dolgoprudny, Moscow Region, 141700, Russia; \\
$^2$ National University of Science and Technology MISIS, 4 Leninsky prosp., Moscow, 119049, Russia; \\
$^3$ Dukhov Research Institute of Automatics (VNIIA), 127055 Moscow, Russia; \\
$^4$ Ginzburg Center for High Temperature Superconductivity and Quantum Materials, P.N. Lebedev Physical Institute of the RAS, 119991, Moscow, Russia; \\
$^5$ National Research Nuclear University MEPhI (Moscow Engineering Physics Institute), 31 Kashirskoye Shosse, 115409 Moscow, Russia; \\
$^6$ Center for Energy Science and Technology, Skolkovo Institute of Science and Technology, Nobel str. 3, 121205 Moscow, Russia; \\
$^7$ HSE University, 101000 Moscow, Russia.
}%

\begin{abstract}
In this work, we report the broad-band ferromagnetic resonance spectroscopy of EuSn$_2$As$_2$ single crystals at different temperatures in combination with magnetization measurements and structural characterization.
We observe conventional collective acoustic resonance mode of the A-type antiferromagnetic spin-flop phase in the Eu sub-lattice, and its transition to the paramagnetic resonance above the ordering temperature.
Furthermore, we observe reproducibly additional well-defined spectral line.
The origin of the additional line remains unclear.
However, its temperature dependence attributes it to magnetism in the Eu sub-lattice.
\end{abstract}

\maketitle

\section{Introduction}

\begin{figure}[!ht]
\begin{center}
\includegraphics[width=0.6\columnwidth]{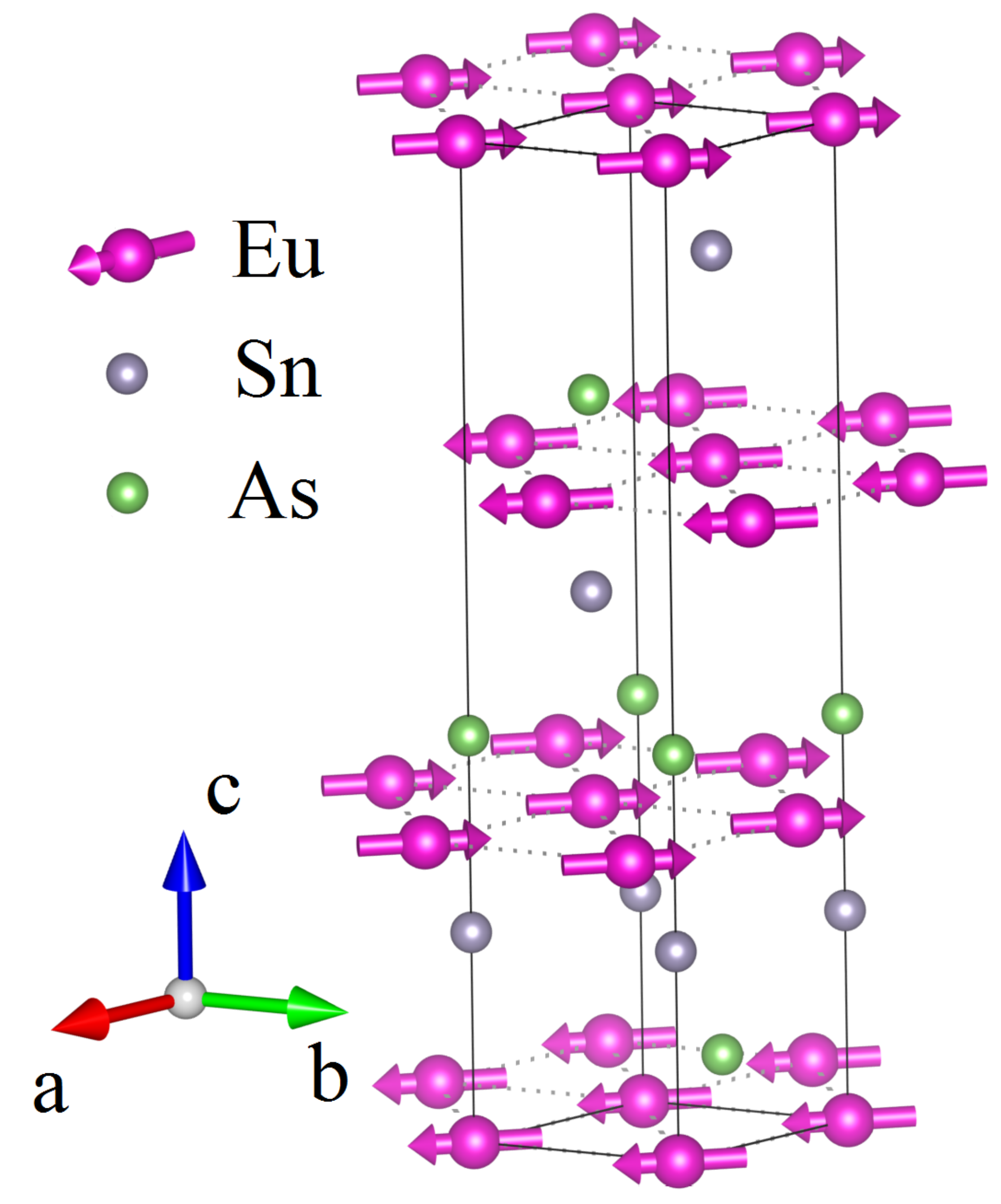}
\caption{Crustal and magnetic structure of EuSn$_2$As$_2$ (made using VESTA software). 
The crystal structure of EuSn$_2$As$_2$ is rhombohedral with the space group R-3m \cite{Arguilla_InChem_4_378}. 
The Eu spin sub-lattice is in the A-type antiferromagnetic state.
The $ab$ crystal plane is the easy magnetic plane.}
\label{cryst}
\end{center}
\end{figure}

In recent years EuSn$_2$As$_2$ compound has been of a high research interest due to its intrinsic topological and magnetic properties \cite{Arguilla_InChem_4_378,Li_PRX_9_041039,Chen_ChPL_37_047201,Zhao_PRL_126_155701,Sun_SciChPMA_64_118211}.
The EuSn$_2$As$_2$ crystallizes in a Bi$_2$Te$_3$-type structure in rhombohedral (R-3m) symmetry and consists of SnAs bilayers sandwiched by hexagonal Eu crystal layers via the van der Waals bonding (see Fig.~\ref{cryst}).
The Eu sub-lattice is in A-type antiferromagnetic order \cite{Pakhira_arXiv}.
The interaction between magnetism and topological surface states in EuSn$_2$As$_2$ induces the dependence of the electronic band structure on magnetic state of Eu sub-lattice \cite{Li_PRX_9_041039}, and can produce many exotic topological effects, including the quantum anomalous Hall effect, Majorana bound states, axion insulator states, and the spin-valve effect \cite{Mong_PRB_81_245209,Qi_PRB_82_184516,Zhang_PRL_122_206401,CHANG_Sci_340_167,DENG_Sci_367_895,Xu_PRL_122_256402,Mogi_NatMat_16_516,Otrokov_PRL_122_107202,Li_PRB_100_121103,Sakuragi_arXiv}. 

In general, electronic and magnetic properties of EuSn$_2$As$_2$ compound are well studied.
Yet, unresolved phenomena remain.
Previous studies demonstrate features in magnetization curves \cite{Li_PRX_9_041039,Pakhira_arXiv,Chen_ChPL_37_047201}, which are argued to be attributed to either paramagnetic impurities \cite{Li_PRX_9_041039,Chen_ChPL_37_047201} or to the presence of low-field in-plane trigonal antiferromagnetic configuration \cite{Pakhira_arXiv}, associated with the 6-fold magneto-crystalline anisotropy within $ab$ crystal planes. 
In this work, we focus on magnetic properties of EuSn$_2$As$_2$ single crystals by combining magnetization measurements and the broad-band ferromagnetic resonance (FMR) spectroscopy.
Ferromagnetic resonance spectroscopy is an ultimate tool for studying exchange properties and magnetic configurations of various magnetic materials \cite{Rezende_JAP_126_151101,Golovchanskiy_arXiv_2108_03847,Golovchanskiy_arXiv_2203_05014,Golovchanskiy_JAP_131_053901,Bogdanov_PRB_75_094425,MacNeill_PRL_123_047204}.
We confirm the A-type antiferromagnetic spin order in the Eu sub-lattice with the ordering temperature about 24~K, and derive anisotropy and exchange characteristics.
Furthermore, we observe reproducibly additional well-defined higher-frequency spectral line at temperatures below the antiferromagnetic ordering temperature.
The origin of the additional line is unclear.
Yet, the characteristic temperature of this line matches the antiferromagnetic ordering temperature of the Eu sub-lattice, which attributes this line unequivocally to magnetism in Eu sub-lattice.
We speculate that this line can be related to resonances of magnetic defects, which may be intrinsic for self-flux-grown EuSn$_2$As$_2$ single crystals.

\section{Experimental details}


The EuSn$_2$As$_2$ single crystals were synthesized from homogeneous SnAs (99.99\% Sn + 99.9999\% As) precursor and elemental Eu (99.95\%) in stoichiometric molar ratio (2:1) using the self-flux method, similar to our previous works \cite{Vlasenko_SUST_33_084009,Eltseva_PhUs_57_827,Vlasenko_JETPLett_107_119}. 
The initial high purity binary compound SnAs (99.99\% Sn + 99.9999\% As) was obtained by solid state reaction technique in quartz ampoule with residual argon atmosphere and was mixed with Eu with a 2:1 molar ratio.
The obtained mixture was placed in an alumina crucible, and sealed in a niobium container with
residual argon pressure. 
The sealed container was heated up in the furnace up to 1250$^\circ$C, and held at this temperature for 24 h to homogenize melting, cooled down to 900$^\circ$C at a rate of 2$^\circ$C/h. 
After that, the container was cooled down to room temperature inside the furnace.

\begin{figure*}[!ht]
\begin{center}
\includegraphics[width=0.99\columnwidth]{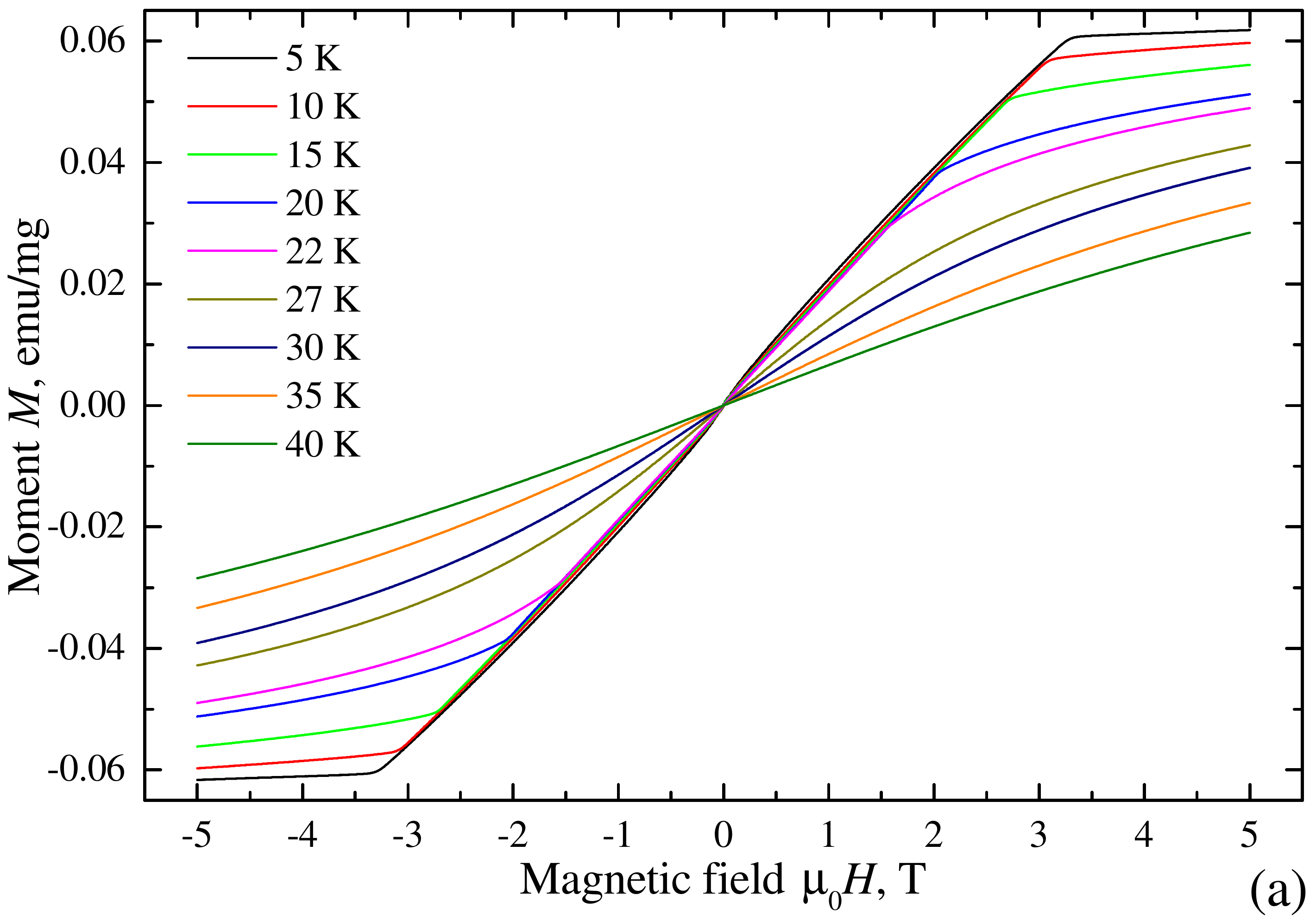}
\includegraphics[width=0.99\columnwidth]{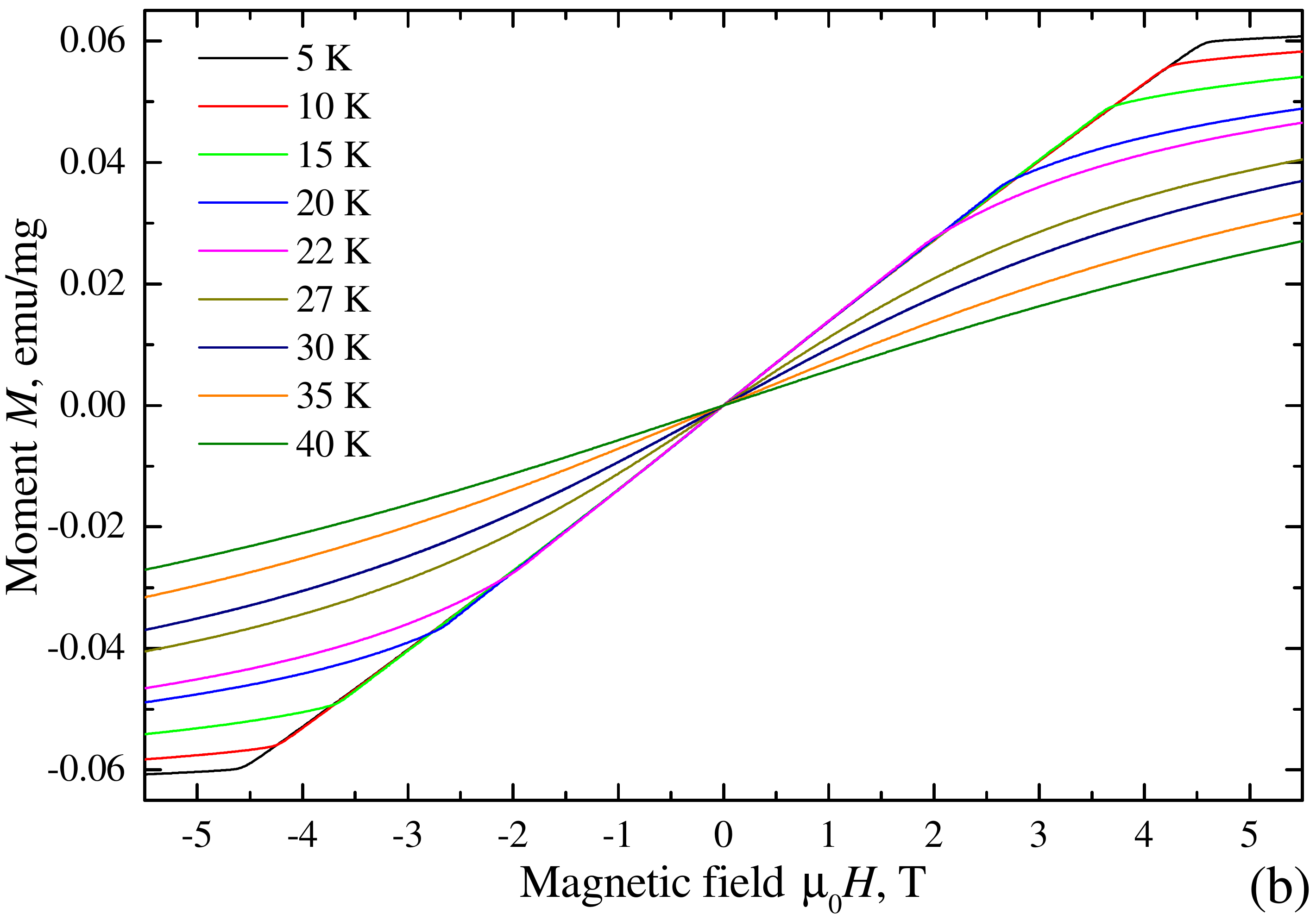}
\includegraphics[width=0.99\columnwidth]{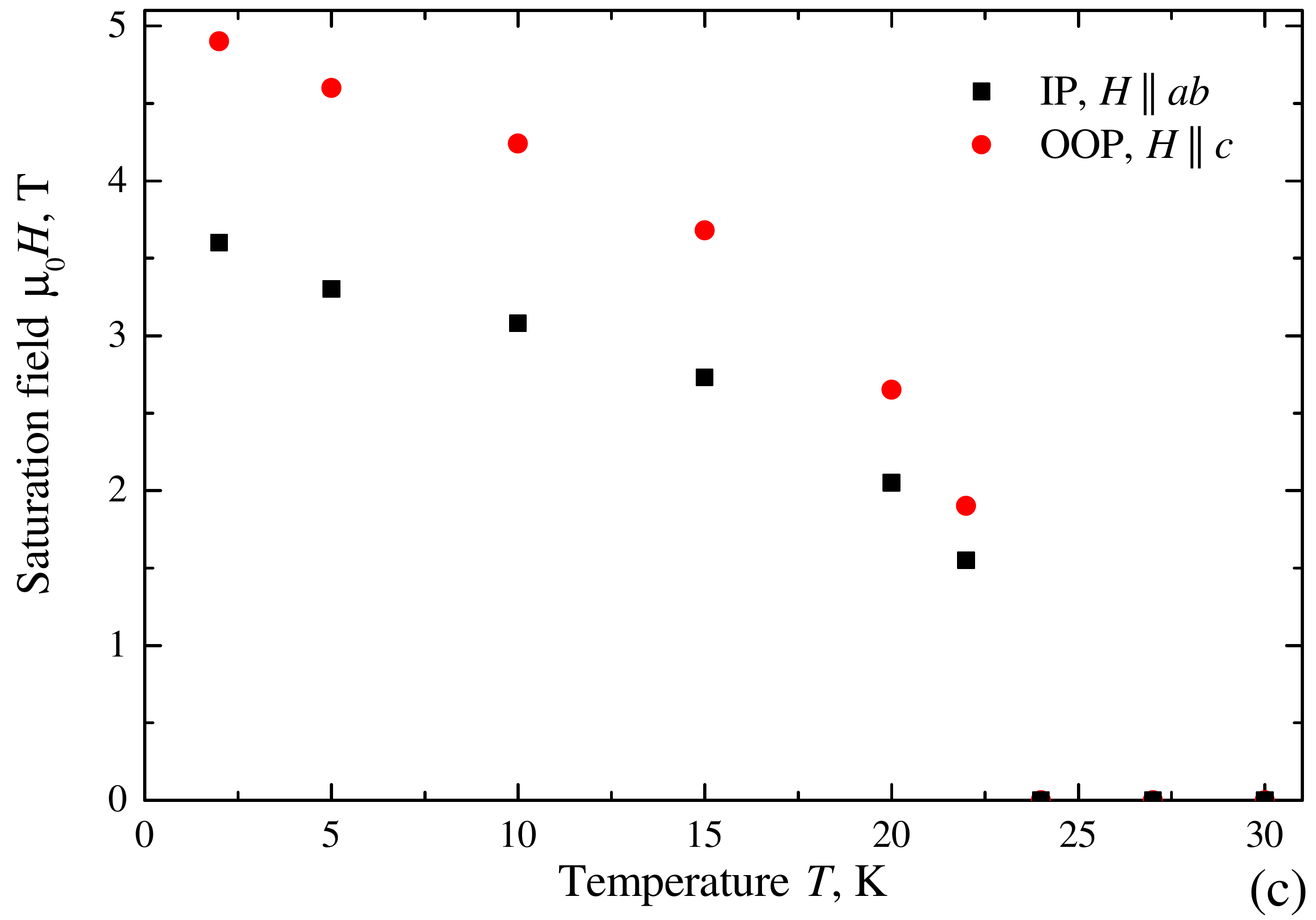}
\includegraphics[width=0.99\columnwidth]{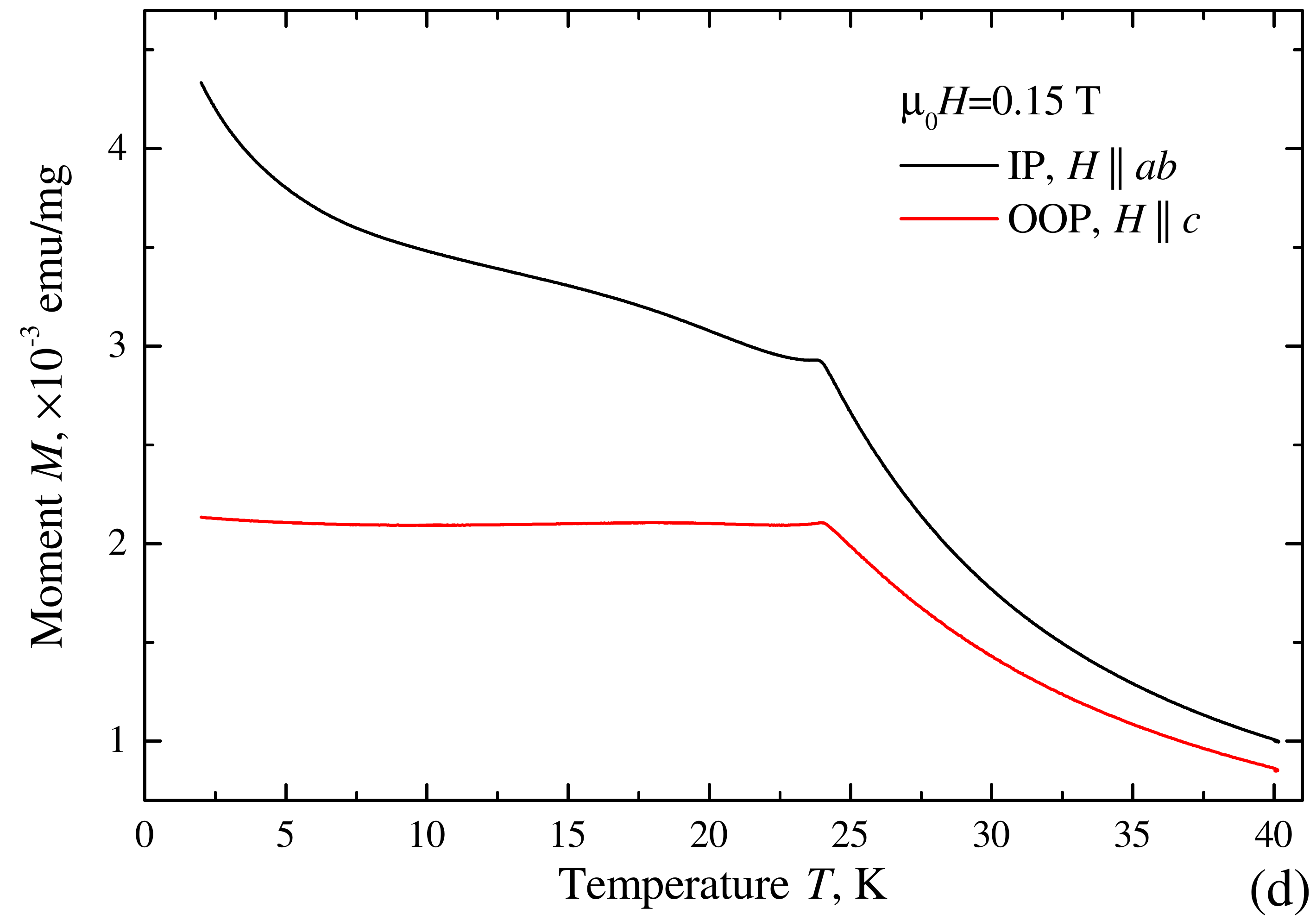}
\caption{a,b) The dependence of magnetization on the magnetic field $M(H)$ at various temperatures in magnetic fields applied in-plane along $ab$ crystal planes (a) and out-of-plane along the $c$ crystal direction.
c) The dependence of the saturation field on temperature for two orientations of the magnetic field.
d) The dependence of magnetization on temperature $M(T)$ at $\mu_0H=0.15$~T for two orientations of the magnetic field.}
\label{M_exp}
\end{center}
\end{figure*}
\begin{figure*}[!ht]
\begin{center}
\includegraphics[width=1\columnwidth]{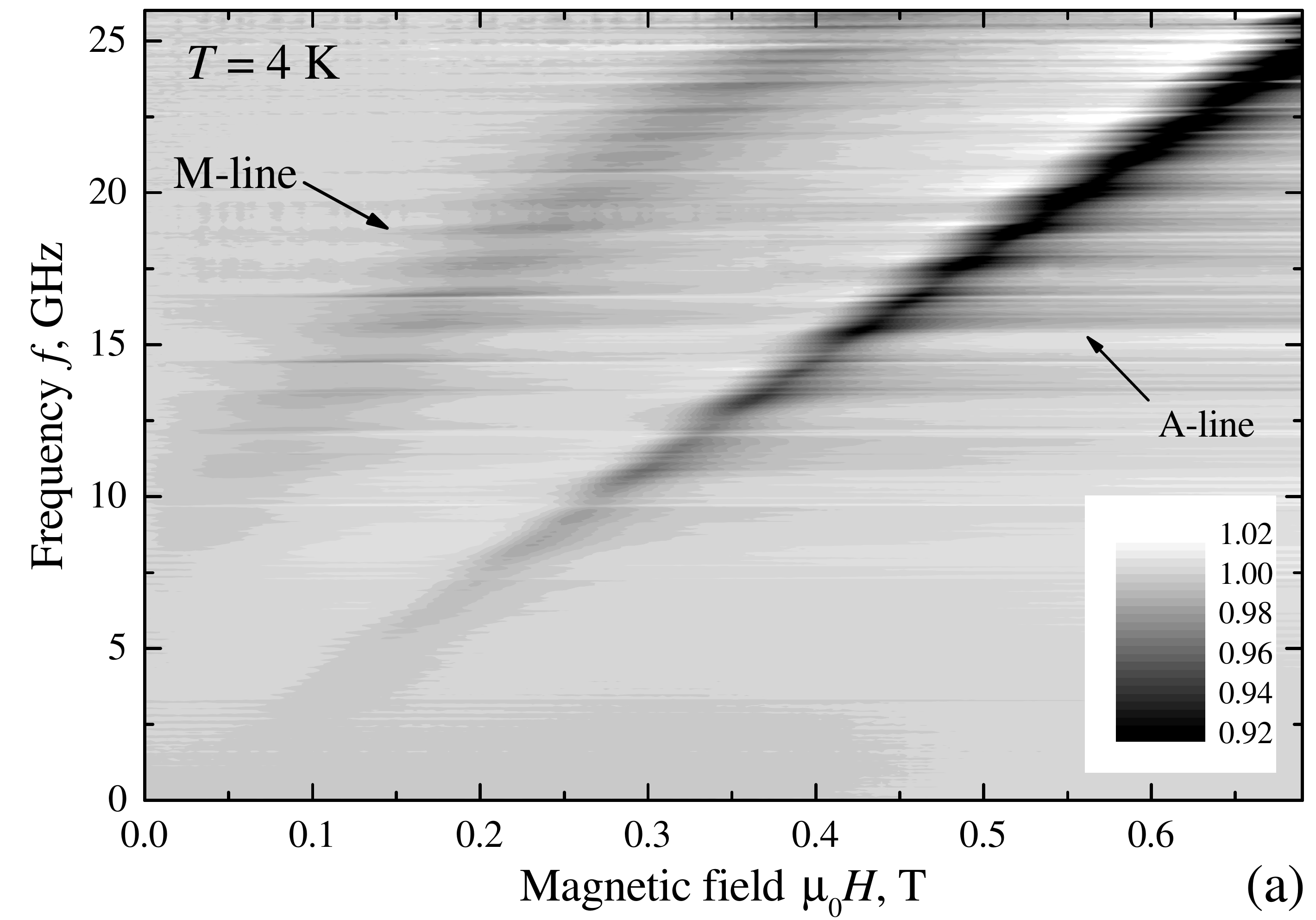}
\includegraphics[width=1\columnwidth]{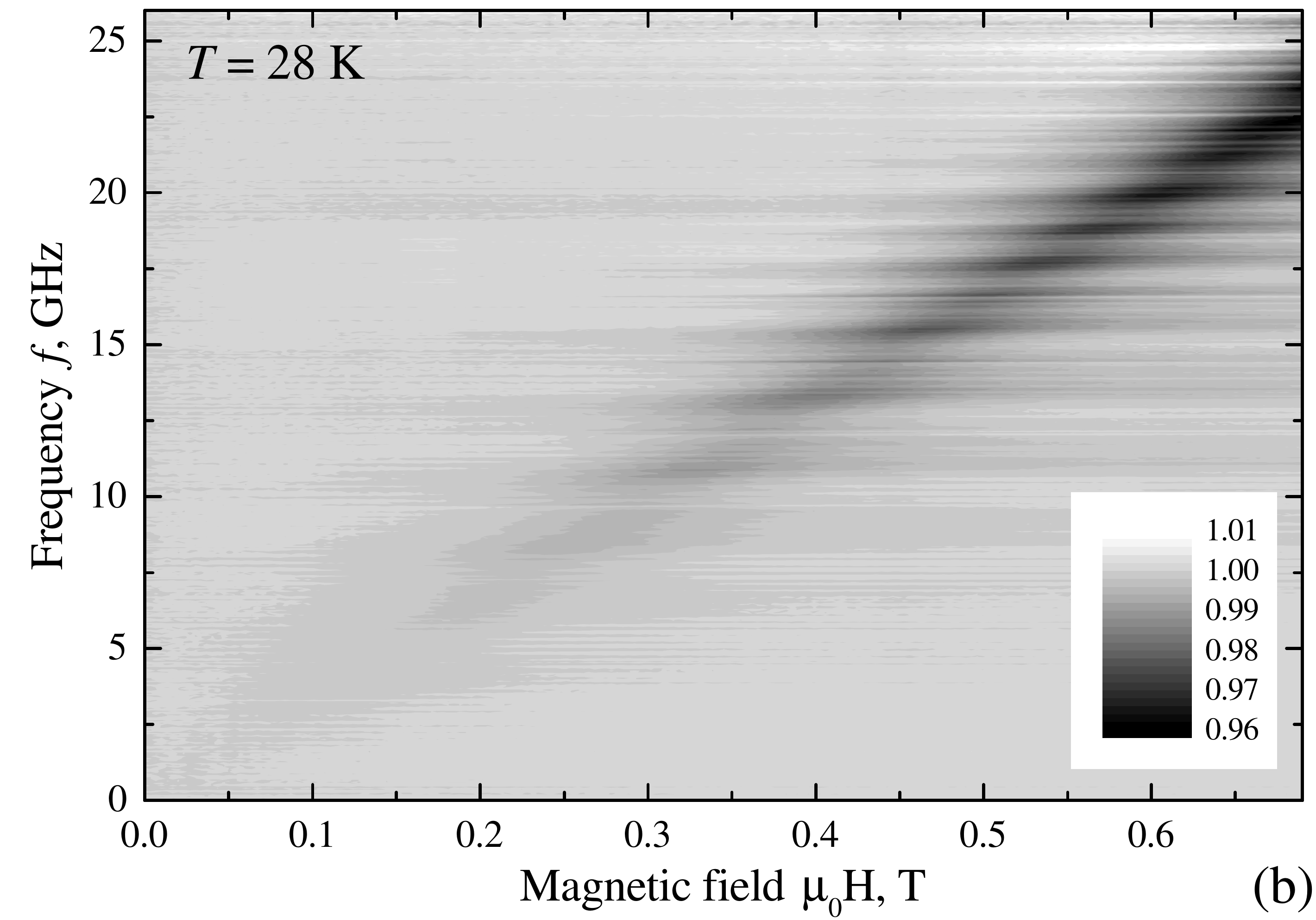}
\includegraphics[width=1\columnwidth]{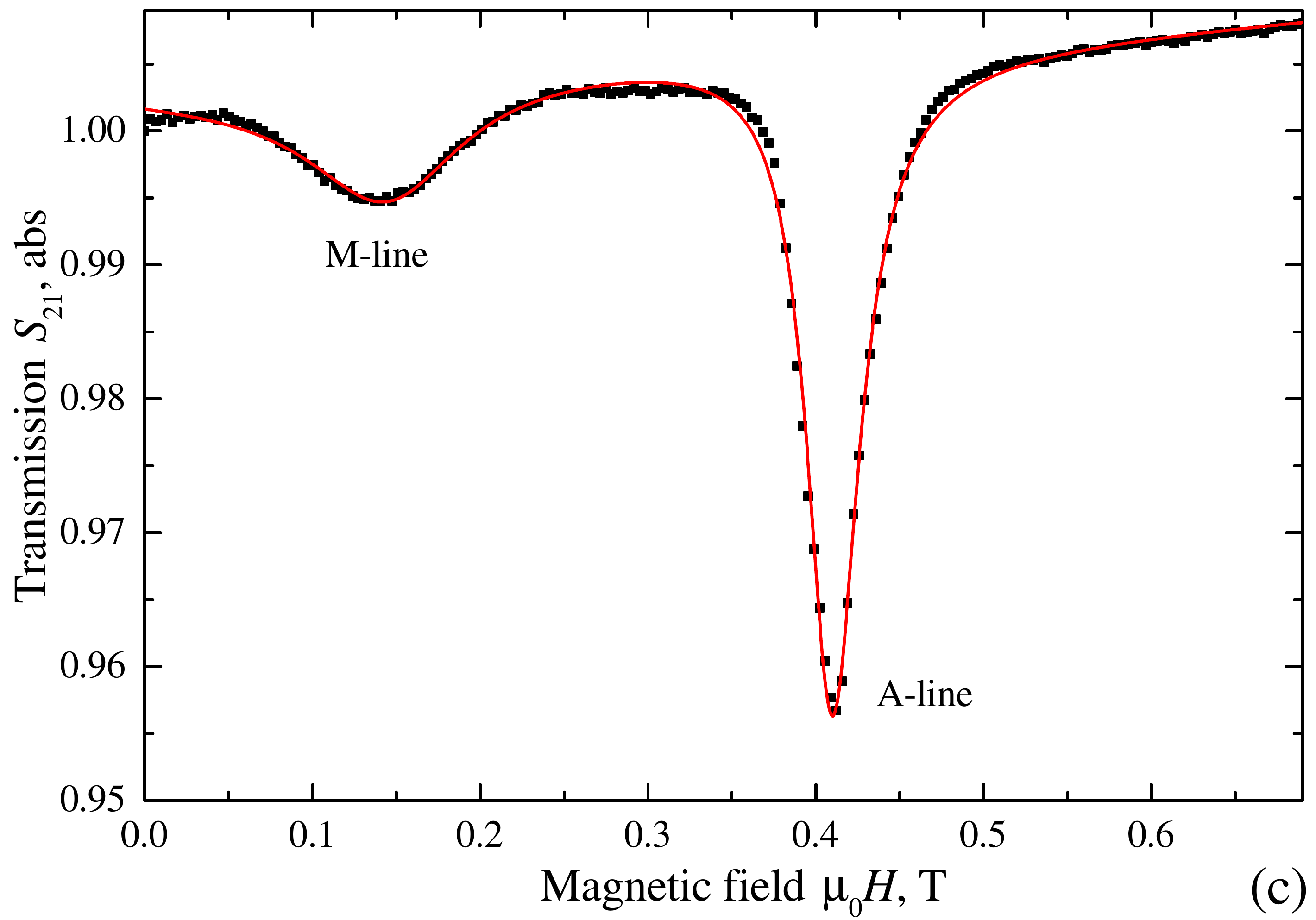}
\includegraphics[width=1\columnwidth]{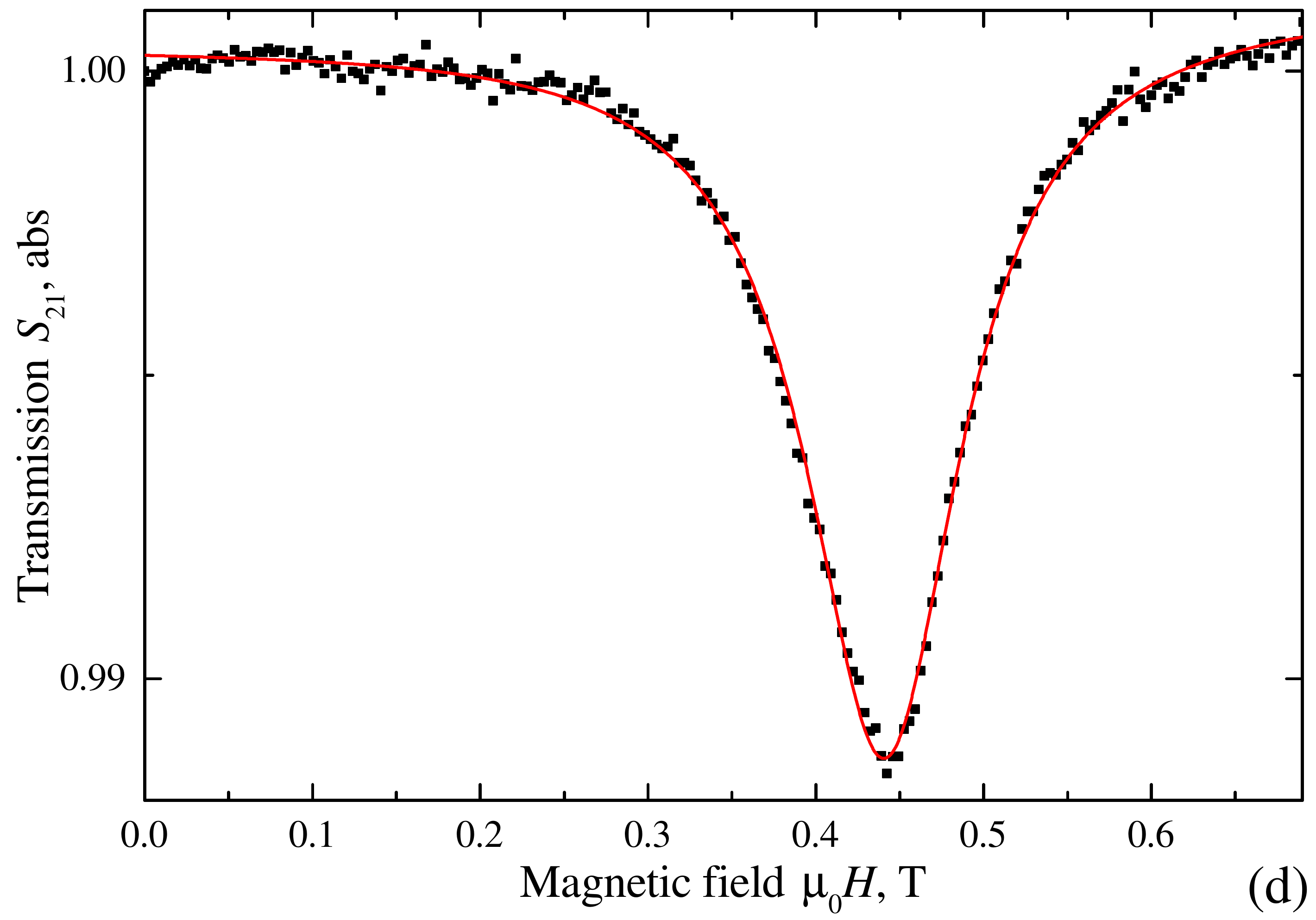}
\caption{a,c) Gray-scale-coded FMR absorption spectra $S_{21}(f,H)$ of EuFe$_2$As$_2$ single crystal measured  at magnetic field applied in-plane along $ab$ crystal planes and at temperatures $T=5$~K (a) and $T=28$~K (c).
b,d) Cross-sections $S_{21}(H)$ of spectra in (a) and (b), respectively, at $f=15$~GHz.
Red curves in (b,d) correspond to fitting of spectral lines with the complex resonance susceptibility\cite{Kalarickal_JAP_99_093909} and linear background.}
\label{FMR_exp}
\end{center}
\end{figure*}

Visually, as-synthesized bulk EuSn$_2$As$_2$ crystals demonstrate well-defined layered structure and their pliability for cleavage and exfoliation along the layering direction only.
The XRD studies confirm alignment of $ab$ crystal planes within the structural layers, and orientation of the $c$ crystal axis across the layers.
The obtained lattice parameters are $a,b=4.189$~\AA  and $c=26.463$~\AA.
The SEM and EDX studies confirm stoichiometric composition (Eu $19.8\pm0.5$~at.\%, Sn $39.7\pm0.2$~at.\%, As $40.5\pm0.3$~at.\%) and homogeneity of synthesized single crystals.
The EBSD studies confirm local crystal orientation of synthesized samples.
Details of XRD and SEM studies can be found in supplementary.
Samples for magnetization measurements and FMR spectroscopy were obtained by cleavage of as-synthesized bulk crystals.
Cleaved EuSn$_2$As$_2$ samples were of a plate shape having few mm in size along $ab$ crystal planes, and from about 100~$\mu$m thick along the $c$ crystal axis, which ensures the well-defined thin-film geometry of samples with determined crystal orientation.


Magnetization measurements of cleaved EuSn$_2$As$_2$ samples were performed using the vibrating-sample magnetometer (VSM) Quantum Design PPMS-9.
The broad-band ferromagnetic resonance (FMR) spectroscopy was performed using the VNA-FMR flip-chip approach \cite{Neudecker_JMMM_307_148,Kalarickal_JAP_99_093909}, by analogy with \cite{Golovchanskiy_arXiv_2108_03847,Golovchanskiy_arXiv_2203_05014}.
Cleaved EuSn$_2$As$_2$ sample was glued on top of the transmission line of coplanar waveguide.
The waveguide with impedance 50~Ohm and the width of the transmission line 0.5~mm is patterned on a Arlon AD1000 copper board and is equipped with SMP rf connectors.
The board with the sample is installed in a brass sample holder.
A thermometer and a heater are attached directly to the holder for precise temperature
control. 
The holder is placed in a home-made superconducting solenoid inside a closed-cycle cryostat (Oxford Instruments Triton, base temperature 1.2 K). 
Magnetic field is applied in-plane along the direction of the waveguide. 
The response of experimental samples is studied by analysing the transmitted microwave signal $S_{21}(f,H)$ with the VNA Rohde \& Schwarz ZVB20.

\section{Results and discussions}


Figure~\ref{M_exp} shows results of magnetization measurements of EuSn$_2$As$_2$ sample for two orientations of the sample in respect to the magnetic field.
The $M(H)$ curves (Fig.~\ref{M_exp}a,b) provide the saturation magnetization $M_s=0.06$~emu/mg.
Also, $M(H)$ curves provide the saturation magnetization $\mu_0H^{ab}_s=3.3$~T when magnetic field is applied along $ab$ planes and $\mu_0H^{c}_s=4.6$~T when magnetic field is applied along the $c$ axis at temperature $T=5$~K. 
At $T=2$~K these values are increased up to $\mu_0H^{ab}_s=3.6$~T and $\mu_0H^{c}_s=4.9$~T.
The dependence of the saturation field on temperature for both orientations is summarised in Fig.~\ref{M_exp}c.
The $M(T)$ curves (Fig.~\ref{M_exp}d) show the magnetic transition at temperature $T_c=24$~K, manifested by the kink, which corresponds to the anti-ferromagnetic spin ordering in Eu sub-lattice of EuSn$_2$As$_2$.
With $M(H)$ data the transition temperature $T_c$ may be noted by the presence of linear-in-field section of $M(H)$ terminated by the kink for all curves at $T<T_c$ and absence of such section with the kink at $T>T_c$ (compare curves $M(H)$ at $T=22$~K and at $T=27$~K in Fig.~\ref{M_exp}a,b).
Both, the saturation fields and the critical temperature, are well consistent with values obtained for EuSn$_2$As$_2$ compound previously \cite{Li_PRX_9_041039,Pakhira_arXiv,Chen_ChPL_37_047201}.

The dependence of magnetization on magnetic field follows the textbook $M(H)$ dependence for the A-type easy-plane antiferromagnet \cite{Gurevich_book,MacNeill_PRL_123_047204}, which is well established for the Eu sub-lattice in EuSn$_2$As$_2$ \cite{Li_PRX_9_041039,Pakhira_arXiv,Chen_ChPL_37_047201}.
At $T\ll T_c$ the free energy expression for the magnetic sublattice is
\begin{equation}
F=2J\vec{e}_{1}\vec{e}_{2}+K_u\sum_{i=1}^{2}e_{iz}^2+\frac{1}{2}M^2_s\sum_{i=1}^{2}\vec{e}_{i}\mathbf{\overline{N}}\vec{e}_{i}-M_s\sum_{i=1}^{2} \vec{e}_i\vec{H},
\label{F_en}
\end{equation}
where $(e_{ix},e_{iy},e_{iz})$ is the unit vector of ferromagnetic moment of the Eu atomic layer in spherical coordinates $\vec{e}=\sin\theta\cos\phi\hat{x}+\sin\theta\sin\phi\hat{y}+\cos\theta\hat{z}$, 
$\hat{z}$ axis is aligned with $c$ crystal axis in Fig.~\ref{cryst}, while 
$[\hat{x},\hat{y}]$ are arbitrary in respect to $[a,b]$,
the first term in Eq.~\ref{F_en} is the isotropic exchange interaction with the exchange energy $J$ between Eu layers,
the second term is the $z$-axis uniaxial anisotropy of individual Eu layers with the anisotropy constant $K_u$, 
the third term is the demagnetizing energy of individual Eu layers with $\mathbf{\overline{N}}=(0, 0, -1)$ being the demagnetizing factor for an infinite thin film,
and the last term is the Zeeman energy with the external field $\vec{H}$.
By treating the saturation process with Eq.~\ref{F_en} as the saturation of the spin-flop phase the saturation fields for in-plane and the out-of-plane orientations are $H^{ab}_s=2H_e=2J/\mu_0M_s$ and $H^{ab}_s=2H_e+M_{eff}=2J/\mu_0M_s+M_s+2K_u/M_s$.
Thus, at $T\ll T_c$ $2J/M_s=3.6$~T and $\mu_0M_{eff}=\mu_0M_s+2K_u/M_s=1.3$~T.
At $H<H^{ab}_s$ or $H<H^c_s$ the dependence of the spin-flop angle on the magnetic field for two sub-lattice configuration are $\cos{\phi}=H/H^{ab}_s$ and $\cos{\theta}=H/H^c_s$, where $\phi$ is the azimuthal angle in $ab$ plane in respect to direction of $H$ when $H$ is in-plane, and $\theta$ is the polar angle in respect to $c$ axis when $H$ is out-of-plane.
These relations imply the linear dependence of magnetization on the magnetic field $M(H)$ up to the saturation field, which is observed in Fig.~\ref{M_exp}a,b.

However, magnetization measurements in Fig.~\ref{M_exp} reveal an extra paramagnetic-like contribution in addition to A-type antiferromagnetic response of Eu layers as follows.
First, $M(H)$ curves demonstrate a slow linear increase of $M$ at $H>H_s$ for both orientations.
Second, some of our samples demonstrate a kink on $M(H)$ curves at small fields at in-plane magnetic field (see supplementary).
Third, $M(T)$ curve shows a progressive increase in $M$ upon decreasing temperature at $T<T_c$ and $H||ab$ (see Fig.~\ref{M_exp}d).
All three features have been observed previously \cite{Li_PRX_9_041039,Pakhira_arXiv,Chen_ChPL_37_047201} and were attributed to either presence of some paramagnetic impurities \cite{Li_PRX_9_041039,Chen_ChPL_37_047201} or to the  presence of low-field in-plane trigonal antiferromagnetic configuration \cite{Pakhira_arXiv}, associated with the 6-fold magneto-crystalline anisotropy. 

\begin{figure*}[!ht]
\begin{center}
\includegraphics[width=0.67\columnwidth]{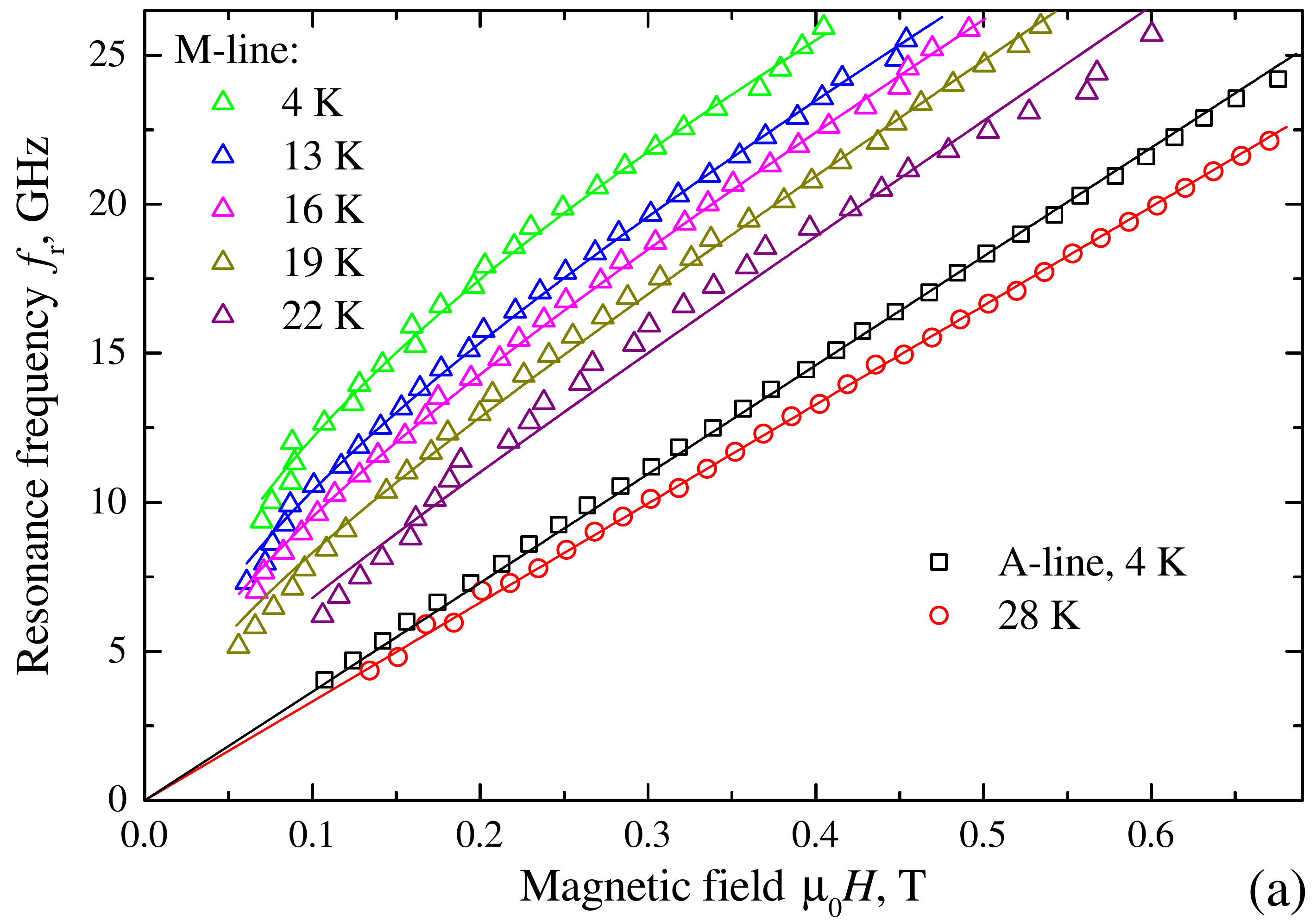}
\includegraphics[width=0.67\columnwidth]{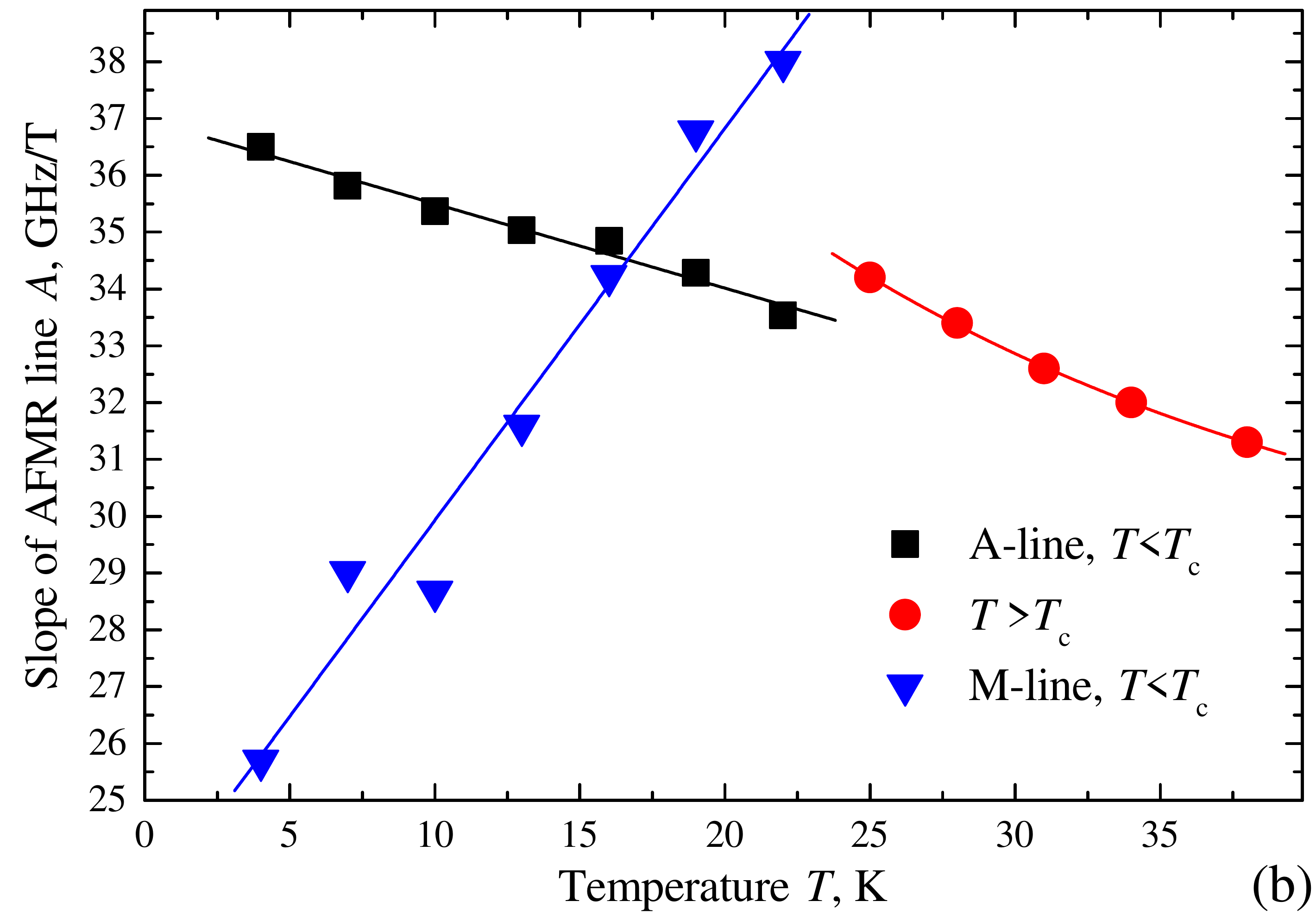}
\includegraphics[width=0.67\columnwidth]{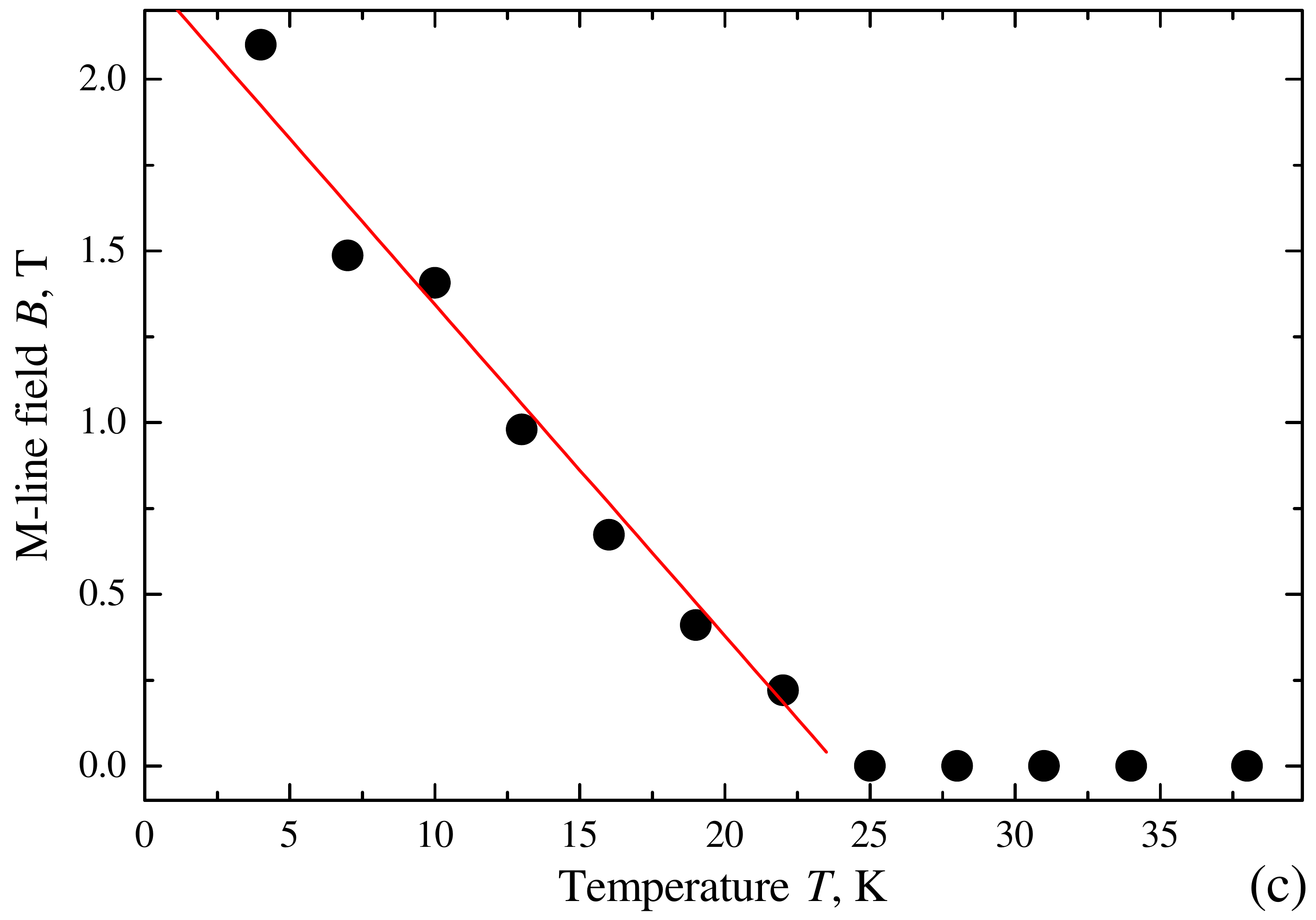}
\caption{a) The dependence of FMR frequency on magnetic field $f_r(H)$ at different temperatures.
Solid lines show modelling of spectral lines with corresponding fitting functions.
b) The dependence of the slope of resonance lines on temperature $A(T)$.
Black and blue solid lines show the linear interpolation of $A(T)$ at $T<24$~K.
Red solid line show an exponential interpolation of $A(T)$ at $T>24$~K. 
c) The dependence of the parameter $B$ of M-line on temperature.
Red solid line shows the linear interpolation of $B(T)$ at $T<25$~K. }
\label{FMR_fit}
\end{center}
\end{figure*}

Figure~\ref{FMR_exp} summarizes results of FMR spectroscopy of the same EuSn$_2$As$_2$ sample as in Fig.~\ref{M_exp}.
At $T<T_c$ (Fig.~\ref{FMR_exp}a) the spectrum consists of two lines:
the linear-in-field stronger spectral line at lower frequencies, refereed to as A-line, and
additional weak spectral line at higher frequencies, refereed to as M-line.
The cross-section of the spectrum at constant $f=15$~GHz is provided in Fig.~\ref{FMR_exp}c.
Both lines were fitted using the complex resonance susceptibility \cite{Kalarickal_JAP_99_093909} and common linear  background as shown with the red curve in Fig.~\ref{FMR_exp}c.
By such modelling of spectral lines at different frequencies the dependence of the resonance frequency on the magnetic field $f_r(H)$ is obtained.
At $T>T_c$ (Fig.~\ref{FMR_exp}b) the spectrum consists of a single paramagnetic resonance line.
The cross-section of the spectrum at constant $f=15$~GHz is provided in Fig.~\ref{FMR_exp}d.
Notice that the resonance field for the strongest line at $T<T_c$ is comparable to the field at $T>T_c$ (compare Figs.~\ref{FMR_exp}b and d). 
Here we emphasize that the demonstrated spectrum is reproducible for other studied samples in terms of the position, the amplitude and the line-width of spectral lines.
Resonance lines $f_r(H)$ at different temperatures are summarized in Fig.~\ref{FMR_fit}a.

At $T<T_c$ (Fig.~\ref{FMR_exp}a) the A-line can be identified as the acoustic antiferromagnetic resonance mode.
Its field dependence $f_r(H)$ can be derived from the free energy expression (Eq.~\ref{F_en}) and follows the linear-in-field expression \cite{Gurevich_book,MacNeill_PRL_123_047204} 
\begin{equation}
2\pi f_r = \mu_0 \gamma\sqrt{1+\frac{M_{eff}}{2H_e}}H,
\label{F_res1}
\end{equation}
while at $T\gtrsim T_c$ the paramagnetic resonance line of a thin-film sample obeys \cite{MacNeill_PRL_123_047204,Golovchanskiy_arXiv_2108_03847}
\begin{equation}
2\pi f_r = \mu_0 \gamma\sqrt{1+\chi(T,H)}H,
\label{F_res2}
\end{equation}
where $\chi(T,H)$ is the residual susceptibility.

%

The evolution of FMR spectrum with temperature is analysed as follows.
In accordance with Eqs.~\ref{F_res1} and \ref{F_res2} the A-line at $T<T_c$ and the spectral line at $T>T_c$ are modelled with the linear function $2\pi f_r=A\mu_0H$ (see Fig.~\ref{FMR_fit}a).
The dependence of the slope on temperature $A(T)$ is shown in Fig.~\ref{FMR_fit}b with squared and circle symbols. 
The magnetic phase transition can be noted by the discontinuity of the slope $A(T)$ at $T\approx T_c$.
By using Eq.~\ref{F_res1}, the extrapolated $A=37$~GHz/T at $T=0$, and saturation fields at 2~K the gyromagnetic ratio is $\gamma/2\pi=31.7$~GHz, which is slightly higher but is comparable with the gyromagnetic ratio of free electrons 28~GHz/T. 

The M-line clearly demonstrates non-linear field dependence and also commencement from $f_r=0$ at $H=0$.
Without implying particular physical meaning, the M-line is fitted with the square root function $2\pi f_r=A\mu_0\sqrt{H(H+B)}$ (see solid lines in Fig.~\ref{FMR_fit}a).
Parameters $A(T)$ and $B(T)$ are shown in Fig.~\ref{FMR_fit}b and c, respectively.
Figure~\ref{FMR_fit}c demonstrates an important finding: 
the resonance process behind the M-line is characterised by the same critical temperature as the A-type anti-ferromagnetic ordering of Eu-sub-lattice.

In general, several possible origins can be proposed for appearance of the additional line.
First, the straightforward possibility of phase inhomogeneity can be considered and is ruled out by reproducible XRD and SEM studies.
Next, the resonance process in a domain structure can be suggested, such as breathers in striped domain structures \cite{Ebels_PRB_63_174437}.
However, EuSn$_2$As$_2$ clearly demonstrates the easy-plane anisotropy.
Moreover, the magnetization per Eu atom is about $8\mu_B$ \cite{Li_PRX_9_041039,Pakhira_arXiv,Chen_ChPL_37_047201}, where $\mu_B$ is the Bohr magneton, which provides the saturation magnetization $\mu_0 M_s\approx 0.52$~T per formula unit in Fig.~\ref{cryst}.
This implies that $c$ axis as the hard magnetocrystalline axis ($2K_u/\mu_0M_s=M_{eff}-M_s>0$), and, therefore, development of the domain structure is not expected.
At last, the spin-wave resonance can be suggested.
In general, magnetic resonance precesses with non-zero wavevector do result in appearance of additional spectral lines in exchange \cite{Kittel_PR_100_1295,Klingler_JPDAP_48_015001,Golovchanskiy_arXiv_2203_05014,Golovchanskiy_JAP_131_053901} or magnetostatic \cite{Golovchanskiy_AdvFuncMater_28_1802375} regimes.
Yet, regardless of particular mechanism or direction of the standing wave any such process is characterised by non-zero resonance frequency at zero field $f_r(H=0)>0$, while the M-line in Figs.~\ref{FMR_exp}a and \ref{FMR_fit}a starts from $f_r=0$ at $H=0$.
Thus, presence of the M-line can not be explained by conventional mechanism. 

As an alternative, we speculate that magnetic defects, i.g., magnetic point defects \cite{Watkins,Dagnelund_APL_102_021910} or stacking-faults, may be the origin of the additional resonance line.
Presence of such defects may have a crucial effect on magnetoresistance transport properties of EuSn$_2$As$_2$.
In fact, the M-line at $T=4$~K follows the Kittel formula for thin films at in-plane magnetic fields $2\pi f_r = \mu_0\gamma\sqrt{H(H+M_{eff})}$, where $\gamma/2\pi=28$~GHz/T and $\mu_0 M_{eff}=1.7$~T. 
This indicates a possible magnetic resonance of individual uncoupled magnetic layers.
The increase in effective magnetization from 1.3~T, defined from magnetization data in Fig.~\ref{M_exp}, up to 1.7~T may be a signature of additional anisotropy induced by a stacking-faults, or of increased saturation magnetization $M_s$ due to substitution impurities. 
Though, no features attributed presence of such phase can be found in magnetization measurements in Fig.~\ref{M_exp}.

\section{Conclusion}

In conclusion, in this work we report the broad-band ferromagnetic resonance spectroscopy study of EuSn$_2$As$_2$2 single crystals in combination with magnetization measurements and structural analysis.
At temperatures above the antiferromagnetic ordering temperature of Eu sub-lattice the spectrum shows a conventional paramagnetic resonance line.
At temperatures below the ordering temperature the spectrum shows the acoustic mode of the A-type antiferromagnetic spin-flop phase.
Also, the spectrum reveals reproducibly additional resonance line. 
The origin of additional line remains unclear, possible conventional explanations fail.
However, the characteristic temperature of this line matches the antiferromagnetic ordering temperature of the Eu sub-lattice, which attributes this line to magnetism in Eu sub-lattice.
As a probable origin we suggest that additional line is a signature of some magnetic defects, which are intrinsic for the self-flux growth process of EuSn$_2$As$_2$2 compound.

\section{Acknowledgments}

This work was supported by the Russian Science Foundation and by the Ministry of Science and Higher Education of the Russian Federation.

\bibliographystyle{apsrev}
\bibliography{A_Bib_EuSnAs}

\end{document}